\documentclass[aps,prd,twocolumn,nofootinbib,showpacs,floats,floatfix,superscriptaddress]{revtex4}

\usepackage{amsfonts}
\usepackage{amsmath}
\usepackage{amssymb}
\usepackage{bm}
\usepackage{dcolumn}
\usepackage{graphicx}
\usepackage{graphics}
\usepackage[latin1]{inputenc}
\usepackage{latexsym}
\usepackage{rotating}
\usepackage{hyperref}
\usepackage[all]{hypcap} 
\usepackage{xspace} 
\usepackage[usenames,dvipsnames]{color}
\usepackage{mathrsfs}
\usepackage{subfigure}
\usepackage{aas_macros}
\usepackage{amsfonts}
\usepackage{booktabs}
\usepackage{siunitx}

\usepackage{makecell}

\usepackage{ulem}
\usepackage{outlines}
\usepackage{enumitem}
\setenumerate[1]{label=\Roman*.}
\setenumerate[2]{label=\Alph*.}
\setenumerate[3]{label=\roman*.}
\setenumerate[4]{label=\alph*.}

\definecolor {darkgreen}{rgb}{0.2,0.7,0.2}

\usepackage{mathrsfs}
\usepackage{aas_macros}
\usepackage{booktabs}
\usepackage{cleveref}

\newcommand{\be}{\begin{eqnarray}}
\newcommand{\ee}{\end{eqnarray}}

\newcommand{\Edd}{{\mbox{\tiny Edd}}}
\newcommand{\obs}{{\mbox{\tiny obs}}}
\newcommand{\pix}{{\mbox{\tiny pix}}}
\newcommand{\rred}{{\mbox{\tiny red}}}
\newcommand{\model}{{\mbox{\tiny mod}}}
\newcommand{\inj}{{\mbox{\tiny inj}}}

\begin{document}

\title{Experimental Relativity with Accretion Disk Observations}

\author{Alejandro C\'ardenas-Avenda\~no} 
\email[Corresponding author: ]{a.cardenasavendano@montana.edu}
\affiliation{eXtreme Gravity Institute, Department of Physics, Montana State University, 59717 Bozeman MT, USA}
\affiliation{Programa de Matem\'atica, Fundaci\'on Universitaria Konrad Lorenz, 110231 Bogot\'a, Colombia}

\author{Jaxen Godfrey} 
\affiliation{eXtreme Gravity Institute, Department of Physics, Montana State University, 59717 Bozeman MT, USA}

\author{Nicol\'as Yunes}
\affiliation{eXtreme Gravity Institute, Department of Physics, Montana State University, 59717 Bozeman MT, USA}

\author{Anne Lohfink}
\affiliation{eXtreme Gravity Institute, Department of Physics, Montana State University, 59717 Bozeman MT, USA}

\date{\today}

\begin{abstract}

Electromagnetic observations have been used over the past decades to understand the nature of black holes and the material around them. Our ability to learn about the fundamental physics relies on our understanding of two key ingredients in the modeling of these electromagnetic observations: the gravity theory that describes the black hole, and the astrophysics that produces the observed radiation. In this work we study our current ability to constrain and detect deviations from General Relativity using the accretion disk spectrum of stellar-mass black holes in binary systems. Our analysis combines relativistic ray-tracing and Markov-Chain Monte-Carlo sampling techniques to determine how well such tests of General Relativity can be carried out in practice. We show that even when a very simple astrophysical model for the accretion disk is assumed \emph{a priori}, the uncertainties and covariances between the parameters of the model and the parameters that control the deformation from General Relativity make any test of General Relativity very challenging with accretion disk spectrum observations. We also discuss the implications of assuming that General Relativity is correct \emph{a priori} on the estimation of parameters of the astrophysical model when the data is not described by Einstein's theory, which can lead to a fundamental systematic bias.
\end{abstract}


\maketitle

\section{Introduction}

Studies of gravitational waves by the LIGO/Virgo collaboration~\cite{ligo2018gwtc}, along with decades of electromagnetic observations~\cite{remillard2006x}, have provided extensive evidence for the existence of stellar-mass black holes (BHs). Are these BHs described by the solutions to Einstein's theory of General Relativity (GR)? 

The no-hair theorems assert that the only axisymmetric, asymptotically flat, electro-vacuum solutions to the Einstein equations are fully described by only three parameters: the mass $M$, the electric charge $Q$ and the angular momentum $S$ of the BH~\cite{Israel:1966nc,1971PhRvL..26..331C,hawking-uniqueness,1975PhRvL..34..905R}. However, in realistic scenarios, black holes should not be significantly charged, as they are expected to be embedded in environments that are rich in gas and plasma and therefore any net charge will be rapidly neutralized~\cite{young1976capture,luminet1998black}.

The \emph{Kerr hypothesis} further states that the exterior spacetime of \emph{astrophysical} BHs is described by the Kerr metric. This hypothesis is a consequence of the no-hair theorem in GR, provided astrophysical BHs are isolated such that the assumptions of the theorem hold. Astrophysical BHs with a time-dependent accretion disk are, of course, not isolated. The assumptions of the theorem then do not hold and the exterior spacetime is not exactly described by the Kerr metric. The accretion disk contributions to the spacetime, however, are very small, and can thus be neglected \cite{Yunes:2011ws,Kocsis:2011dr}. Therefore, observational tests of the Kerr hypothesis can be thought of as tests of GR, without additional fields, as GR is the only assumption left to break in the no-hair theorems.

Over the past decades, several authors have proposed various ways to probe the spacetime geometry of BH candidates (see Refs.~\cite{Psaltis:2008bb, Bambi:2015kza} and references therein for tests with electromagnetic observations, and Ref.~\cite{Yunes:2013dva} and references therein for tests with gravitational waves, and for a general review see Ref.~\cite{Will:2014kxa}).  In particular, one can parametrize a potential deviation from the Kerr metric in terms of a parameter, or a family of parameters. If an observation of a compact object then yields a nonzero deviation, this measurement implies that the no-hair theorems are violated~\cite{yagi2016black}, signaling a deviation from GR (provided the other assumptions of the theorems are sufficiently satisfied within observational uncertainties).  Parametrized deviations from the Schwarzschild or Kerr metrics are commonly referred to as \emph{bumpy} BH metrics.

The construction of bumpy BH spacetimes with arbitrary multipole moments was first investigated by Ryan~\cite{ryan1995gravitational}, then revisited by Collins and Hughes~\cite{collins2004towards}, shortly after by Glampedakis and Babak~\cite{glampedakis2006mapping}, and later by Vigeland and Hughes~\cite{Vigeland:2009pr}. These studies, however, required the bumpy multipoles to satisfy Einstein's equations, which then introduced naked singularities in the spacetime~\cite{collins2004towards}. In an effort to avoid these, recent work has lifted the requirement that the multipoles satisfy the Einstein equations, yielding bumpy BH metrics that could sometimes be mapped to BH solutions in modified theories of gravity~\cite{vigeland2011bumpy, Johannsen:2011dh, cardoso2014generic}. The most recent incarnation of this idea by Konoplya, Rezzolla and Zhidenko~\cite{Rezzolla:2014mua, Konoplya:2016jvv} proposes a parametrization of BH bumps with continued fractions. These bumpy BH metrics have been used to carry out some tests of GR using electromagnetic observations~\cite{bambi2011constraining, vincent2013testing, kong2014constraints, cardenas2016testing, Ayzenberg:2017ufk, bambi2017testing,choudhury2017testing,banerjee2017excavating,niedzwiecki2018lamppost, kammoun2018testing, taylor2018exploring}.

In a recent review~\cite{krawczynski2018difficulties}, Krawczynski pointed out qualitatively that astrophysical uncertainties can be so large that they may prevent us from distinguishing between Kerr and non-Kerr metrics. In general, there are many sources of uncertainty when carrying out and analyzing electromagnetic observations. In addition to statistical uncertainties in the analysis of the data, there are typically also systematic uncertainties related to calibration and to the astrophysical modeling~\cite{maccarone2013black}. The latter is particularly problematic, as one must choose a particular physical model that is not necessarily complete, either because of our limited knowledge of the astrophysics, or because one might be forced to ignore important physical processes that are too difficult to incorporate. For example, magnetic fields, the radiation pressure, thickness of the disk, effects of a warm absorber are often ignored in accretion disk models in order to make computations more manageable~\cite{abramowicz2013foundations, davis2006testing}. 

Bearing in mind these limitations, an important question arises: Can one truly use bumpy BH metrics to test GR with electromagnetic observations? From a general perspective, the answer could be ``no" for several reasons. Bumpy BH metrics could contain physical pathologies that would rule them out from the beginning, such as naked singularities or closed time-like curves outside the event horizon. But even if one uses more physically-reasonable bumpy metrics, degeneracies between the parameters that characterize the deformations and those of the astrophysical model may make these tests uninformative. Whether this is the case or not is what we study in this paper using Bayesian statistical theory. 

As a proof-of-concept, we work here with the Rezzolla-Zhidenko (RZ) metric~\cite{Rezzolla:2014mua}, which represents a physically-reasonable bumpy BH metric without spin~\cite{krawczynski2018difficulties}. The RZ metric includes the Schwarzschild metric in the limit as the bumpy deformation parameters vanish, and it can also be mapped to a wide range of BH solutions in modified theories of gravity. Therefore, in principle, if observational constraints on the bumpy parameters are possible, these could be translated into constraints on the coupling parameters of modified theories. 

Given the RZ metric, we then construct spectra from a geometrically thin and optically thick accretion disk characterized by certain disk model parameters. We begin by prescribing the temperature of the disk using a Novikov-Thorne model~\cite{novikov1973astrophysics} and assuming the emission to be blackbody-like locally. This temperature profile is then used by the relativistic ray-tracing code GYOTO~\cite{vincent2011gyoto}, which solves for the motion of a photon ray from the image screen to the accretion disk (integrating backwards in time). We ray trace over $10,000$ photon rays to generate the accretion disk spectrum on a $129 \times 129$ pixel screen.

With this machinery in hand, we then carry out several different studies. We first create a synthetic signal assuming GR for a set of injected disk model parameters, and we attempt to recover it with accretion disk spectra generated with a non-GR model. The latter is determined by 5 parameters: the accretion rate, the BH mass, the distance to the observer, the inclination angle and a single bumpy parameter. We carry out a Markov-Chain Monte-Carlo (MCMC) exploration of the likelihood surface to find the marginalized posterior distribution for each parameter in the non-GR model. We find that the marginalized posterior of the bumpy parameter is close to the flat prior we chose on this parameter, implying the observation was uninformative for this parameter. This is true whether we simulate synthetic signals from current telescopes or from future telescopes that are expected to be able to measure disk parameters one order of magnitude more accurately. We also find that if one freezes the disk parameters to the injected values and only varies over the bumpy parameter, one is typically misled to believe that a strong constraint is possible. The reason is clearly that by freezing disk parameters one is ignoring strong covariances between the bumpy parameter and disk model parameters (in particular the accretion rate and the inclination angle).   

Our second study is aimed at determining whether current parameter estimation of disk model parameters assuming GR is correct could be systematically biased by this \emph{a priori} assumption. We thus create a synthetic signal assuming a bumpy BH metric with a fixed bumpy deformation parameter and a set of injected disk model parameters, and we attempt to recover it with accretion disk spectra generated within GR. We find that for signals that do not deviate much from GR, the amount of fundamental bias in the recovered disk model parameters is minimal and within the statistical uncertainties. Only once the bumpy deformation becomes strong enough -typically well above already ruled out values in particular modified theories- does fundamental bias systematically affect parameter estimation of the disk model parameters. 

Finally, we study whether a signal generated from a non-GR BH metric could be distinguished from one produced in GR. We thus create again a synthetic signal in a bumpy BH metric (with fixed deformation parameter and disk model parameters), and we attempt to recover it with bumpy BH accretion disk spectra. As in the fundamental bias case, we discover that if the injected deformation is large enough, then the marginalized posterior distribution of the bumpy parameter peaks away from zero, signaling a GR deviation. However, for smaller values of the injected deformation, the modified spectrum is not sufficiently different from those produced in GR to break the strong degeneracies with the disk model parameters. 

Our study therefore suggests that electromagnetic tests of GR with accretion disk observations that are competitive with other tests, such as those carried out in the Solar System~\cite{Will:2014kxa}, with binary pulsars~\cite{Stairs:2003eg} or with gravitational waves~\cite{Yunes:2013dva}, are, at the very least, very challenging with current and future telescopes. In part, this is due to degeneracies between disk model parameters and bumpy parameters that tend to overwhelm the likelihood and prevent constraints on the latter; we have verified that such a conclusion is robust against different injected parameters of the synthetic signal. But in part, these challenges are also because of other fundamental limitations in the spectral analysis that we did not include in our studies, such as calibration uncertainties, uncertainties in the overall modeling of the disk (such as through the inclusion of thickness~\cite{taylor2018exploring}), and exacerbated degeneracies when the disk model also includes spin~\cite{kammoun2018testing} values. The inclusion of such limitations should make our conclusions even stronger.  

The remainder of this paper presents the details of the results summarized above and it is organized as follows:
Section~\ref{AbBH} briefly reviews the RZ metric; Sec.~\ref{Model} summarizes the accretion disk model used and the general expressions tailored for this particular background;
Sec.~\ref{MCMC} describes the MCMC methods used to explore the likelihood surface;
Sec.~\ref{Simulations} shows the results of the different type of simulations performed;
Sec.~\ref{Conclusions} concludes and points to future work. 
Throughout the paper, we mostly use geometric units in which $G=1=c$, and the $(-,+,+,+)$ metric signature. 
Commas in index lists will stand for partial derivatives. 

\section{A bumpy BH metric}
\label{AbBH}
In this section, we establish the notation by briefly summarizing the parametric solutions proposed in Ref.~\cite{Rezzolla:2014mua} to describe bumpy BH metrics. For simplicity, we will here keep only a single bumpy deformation parameter; we could retain more, but their inclusion would only increase the level of degeneracy and thus hinder any constraints even further. 

The starting point of the bumpy BH metrics of Ref.~\cite{Rezzolla:2014mua} is a spherically-symmetric and static line element, which in spherical polar coordinates $\left(t,r,\theta,\phi\right)$ can be written as
\begin{equation}
ds^{2}=-N^{2}\left(r\right)dt^{2}+\frac{B^{2}\left(r\right)}{N^{2}\left(r\right)}dr^{2}+r^{2}d\Omega^{2},\label{eq:linelement}
\end{equation}
where $d\Omega^{2}\equiv d\theta^{2}+\sin^{2}\theta \, d\phi^{2}$ is the line element on the two-sphere. If such a line element is to represent the exterior spacetime of a BH, then it must contain an event horizon, i.e., a null hypersurface generated by null geodesics with vanishing expansion, whose location $r=r_{0}>0$ is given by $N(r_{0})=0.$ This line element can be recast in terms of a compactified radial coordinate 
\begin{equation}
x\equiv1-\frac{r_{0}}{r},
\end{equation}
so that $x=0$ corresponds to the location of the event horizon, while $x=1$ corresponds to spatial infinity. 

With this at hand, we must now choose a parameterization of the metric functions $N(x)$ and $B(x)$. The first one can be expressed as
\begin{equation}
N^{2}(x)=xA(x)
\end{equation}
for some other function $A(x)>0$ that Ref.~\cite{Rezzolla:2014mua} chooses to write as
\begin{equation}
A(x)=1-\epsilon(1-x)+(a_{0}-\epsilon)(1-x)^{2} + {\cal{O}}[(1 - x)^{3}]\,,
\end{equation}
and similarly
\begin{equation}
B(x)=1+b_{0}(1-x) + {\cal{O}}[(1 - x)^{2}]\,.
\end{equation}
The terms proportional to $(1-x)^{3}$ in $A(x)$ and $(1-x)^{2}$ in $B(x)$ are identically zero when one sets all additional bumpy parameters to zero, as we do in this paper. The line element is then characterized in terms of the constant bumpy parameters $\epsilon$, $a_{0}$, and $b_{0}$, which characterize the magnitude of the non-Schwarzschild deformation.  
 
The physical meaning of these bumpy parameters can be inferred by studying the post-Minkowskian limit  of the line element. An expansion of the metric about spatial infinity reveals that the bumpy parameters are related to the parameterized post-Newtonian (PPN) parameters $(\beta,\gamma)$ via~\cite{Will:2014kxa}
\begin{align}
\beta & =  1+\frac{2\left[a_{0}+b_{0}(1+\epsilon)\right]}{(1+\epsilon)^{2}}
\\
\gamma & =  1+\frac{2b_{0}}{1+\epsilon}\,, 
\end{align}
while the horizon location is related to $\epsilon$ via
\begin{align}
1+\epsilon & =  \frac{2M}{r_{0}}  \label{epsilonRZ}\,,
\end{align}
where $M$ is the ADM mass of the spacetime. From the current observational constraints~\cite{Will:2014kxa} of $\beta$ and $\gamma$, the lowest order bumpy parameters are constrained to $a_{0} \sim10^{-4} \sim b_{0},$ and therefore, in this paper, we will set $a_{0}$ and $b_{0}$ to zero and consider only deviations in the metric due to a non-vanishing $\epsilon$, which has a clear physical meaning in terms of corrections to the event horizon location.

\section{Accretion disk model}
\label{Model}
The spectral energy distribution of X-ray binaries is commonly dominated by a thermal component and a power-law tail. The thermal spectral component, the subject of this work, is believed to be quasi black body emission from the accretion disk that peaks around $kT \sim 0.1-1\, \text{keV}$ in case of stellar-mass black holes.The power-law tail is believed to originate from the Compton up-scattering of thermal photons by a hot electron corona. In this work, we focus on the \emph{thermal component} of the X-ray spectrum of a stellar-mass BH with a low-mass companion. For this particular type of systems, the X-ray region is the most informative feature of the spectrum about the spacetime metric, given the fact that the power drops away from the peak by orders of magnitude and that longer wavelength emission is produced at larger radii, where GR or any deviations thereof are unimportant. The accretion disk is modeled by the Novikov-Throne model~\cite{novikov1973astrophysics}, i.e.~a geometrically thin ($h/r\ll1$, where $h$ is the semi-thickness of the disc at a radial coordinate $r$) and optically thick disk (the photon mean free path $l=(n\sigma_{phot})^{-1}\ll h$, where $\sigma_{phot}$ is the photon scattering cross-section in the disk medium and $n$ is the number density of scattering particles in the disk). 

The disk spectrum depends on the temperature profile of the disk $T(r)$, which in turn determines on the radial flux $\mathcal{F}(r)$ via the Stefan-Boltzmann law $\mathcal{F}(r)=\sigma T^{4}(r)$, where $\sigma$ is the Stefan-Boltzmann constant. The time-averaged energy flux emitted from the surface of a geometrically thin and optically thick accretion disc with material in circular rotation is given by~\cite{vincent2013testing} 
\begin{equation}
\mathcal{F}(r)=\frac{\dot{M}}{4\pi}\frac{1}{\sqrt{-g_{rr}g_{tt}g_{\phi \phi}}}\frac{-\Omega_{,r}}{(E-\Omega L)^{2}}\int_{r_{\mathrm{in}}}^{r}(E-\Omega L)L_{,r'}dr'
\label{eq:Fr}
\end{equation}
in spherical polar coordinates, where $g_{rr}$, $g_{tt}$ and $g_{\phi \phi}$ are the $(r, r)$, $(t, t)$ and $(\phi,\phi)$ components of the metric. The inner edge of the disk  $r_{\mathrm{in}}$ is assumed here to coincide with the innermost stable circular orbit (ISCO),  and the disc is assumed to extend to at least $r=300M$.  The specific energy $E$, the ($z$-component of the) specific angular momentum $L$, and the orbital frequency $\Omega$ are found from the metric, assuming the disk is in the equatorial plane ($\theta=\pi/2$) via
\begin{align}
\Omega &= \sqrt{\frac{N'(r)N(r)}{r}},
\label{eq:OmegaRZ}
\\
E&=\sqrt{\frac{N^{3}(r)}{N(r)-rN'(r)}},
\label{eq:EnergyRZ}
\\
L&=\sqrt{\frac{r^{3}N'(r)}{N(r)-rN'(r)}}.
\label{eq:AngularLRZ}
\end{align}

The location of the ISCO $r^{*}$ can be determined by considering the (timelike) geodesic motion of a massive particle. This motion reduces to that of a particle in a  one-dimensional effective potential
\begin{equation}
V_{\mathrm{eff}}(r)=\frac{E^{2}}{N^{2}(r)}-\frac{L^{2}}{r^{2}}-1.
\label{eq:EffV}
\end{equation}
A circular orbit then satisfies the condition 
\begin{equation}
V_{e\mathrm{ff}}(r)=0=V'_{\mathrm{eff}}(r).
\label{eq:Cond1}
\end{equation}
while the innermost stable orbit additionally satisfies the condition
\begin{equation}
V''_{\mathrm{eff}}(r^{*})=0.
\label{eq:Cond2}
\end{equation}
Combining~\eqref{eq:EffV}-\eqref{eq:Cond2}, the ISCO radius is
thus the real non-zero root of
\begin{equation}
3N(r^{*})N'(r^{*})-3r^{*}N(r^{*})N''(r^{*})=0,
\label{eq:ISCOcond}
\end{equation}
which can be solved numerically. 

For BHs, the Eddington luminosity sets a theoretical maximum luminosity
and the Eddington accretion rate is~\cite{bambi2017black} 
\begin{equation}
\dot{M}_{\Edd} \approx 10^{18} \left(\frac{0.1}{\eta_{r}}\right) \left(\frac{M}{M_{\odot}}\right)\, [\mathrm{g} \, \mathrm{s^{-1}}]
\label{eq:Mdot}
\end{equation}
where $\eta_{r}$ is the radiative efficiency, and we have assumed
that $L_{\Edd}=\eta_{r}\dot{M}_{\Edd}c^{2}$. For a BH with $M=10M_{\odot}$
accreting at $\sim10\%$ of the Eddington limit, the mass accretion
rate is $\sim10^{18}\,[\mathrm{g} \, \mathrm{s^{-1}}]$. 

The temperature profile has to be numerically evaluated on a radial grid (as Eq.~(\ref{eq:Fr}) cannot be solved analytically) and provided to the relativistic ray-tracing code GYOTO~\cite{vincent2011gyoto} as an input. GYOTO then solves for the motion of photon rays from the image screen, located a distance $r_{\obs}$ at an inclination angle $i$, to the accretion disk (integrating backwards in time). Once a map of the Planck function $B_{\nu}$ is computed on the screen, the observed flux is then given by~\cite{vincent2013testing} 
\begin{equation}
F_{\nu,\obs}=\sum_{\pix} B_{\nu}\!\left(T_{\pix}\right) \; \cos\theta\frac{\Delta\Omega}{N_{\pix}},
\label{eq:Mdot}
\end{equation}
where the sum is performed over $N_{\pix}=129 \times 129$ pixels, $\theta$ is the angle between the normal of the screen and the current pixel direction,  $\Delta\Omega=\pi \text{FOV}^2/r_{\obs}^2$ is the solid angle subtended by the screen and $\text{FOV}$ is the field-of-view.

\begin{figure*}
\includegraphics[width=.4\textwidth]{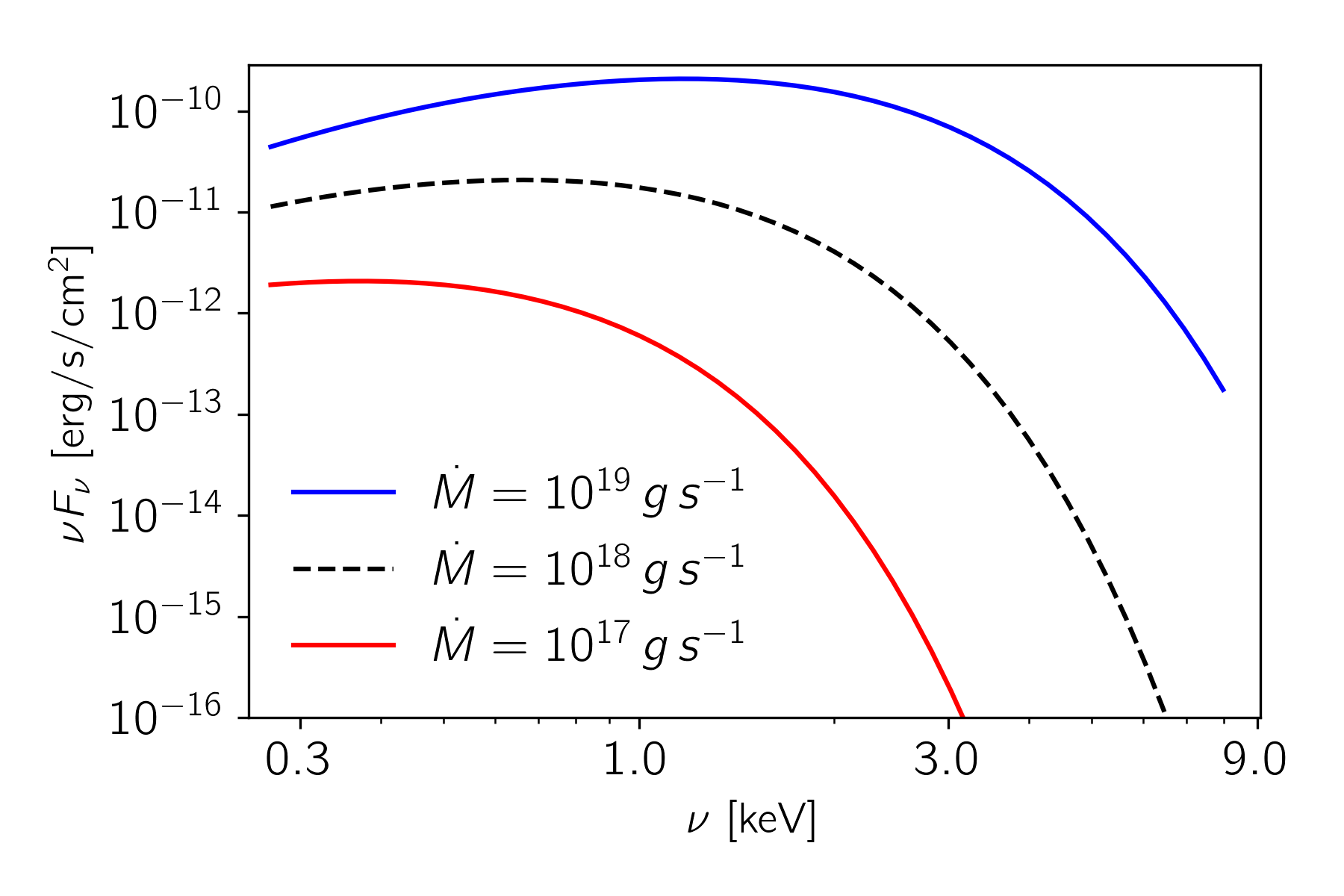} 
\includegraphics[width=.4\textwidth]{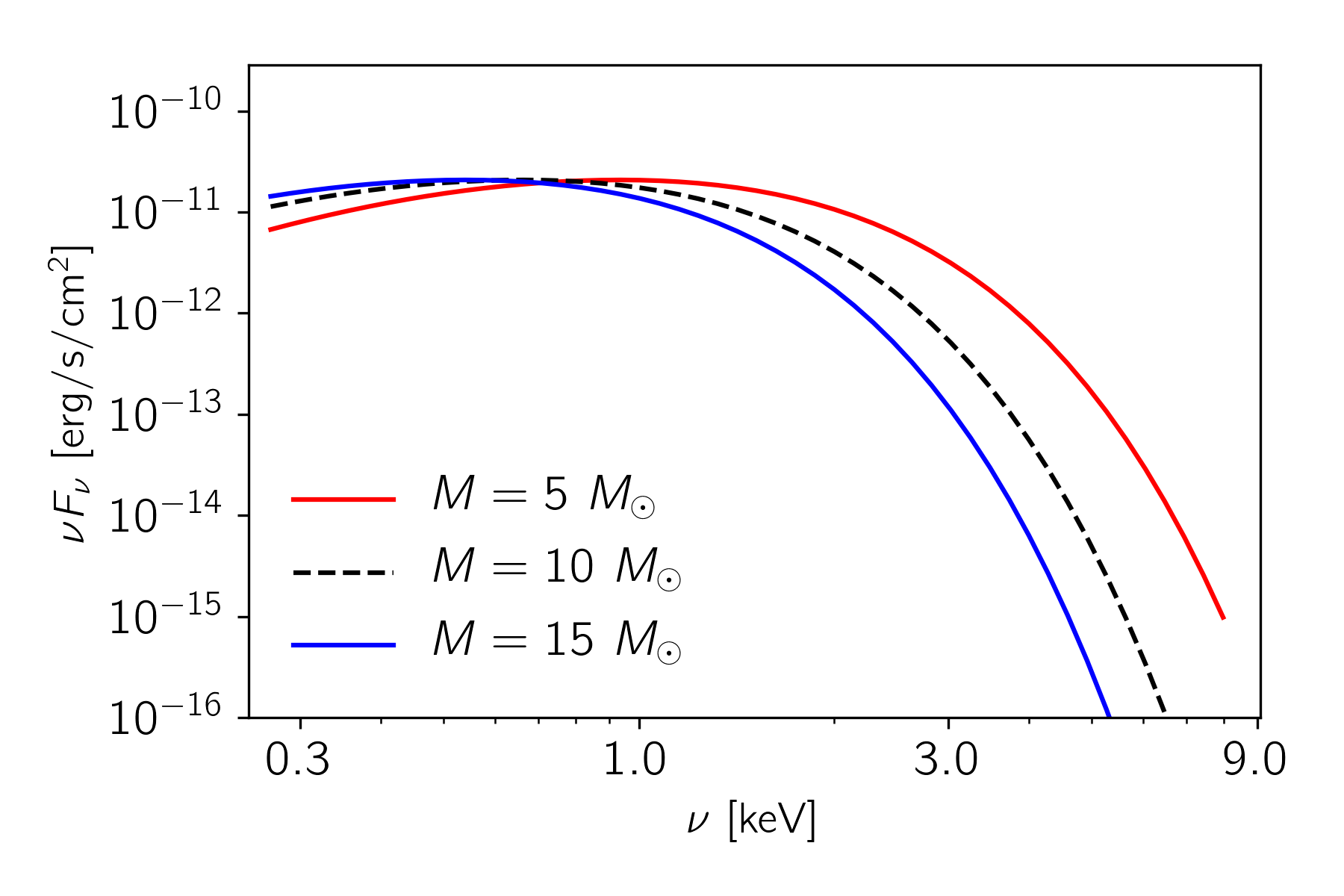} \\
\includegraphics[width=.4\textwidth]{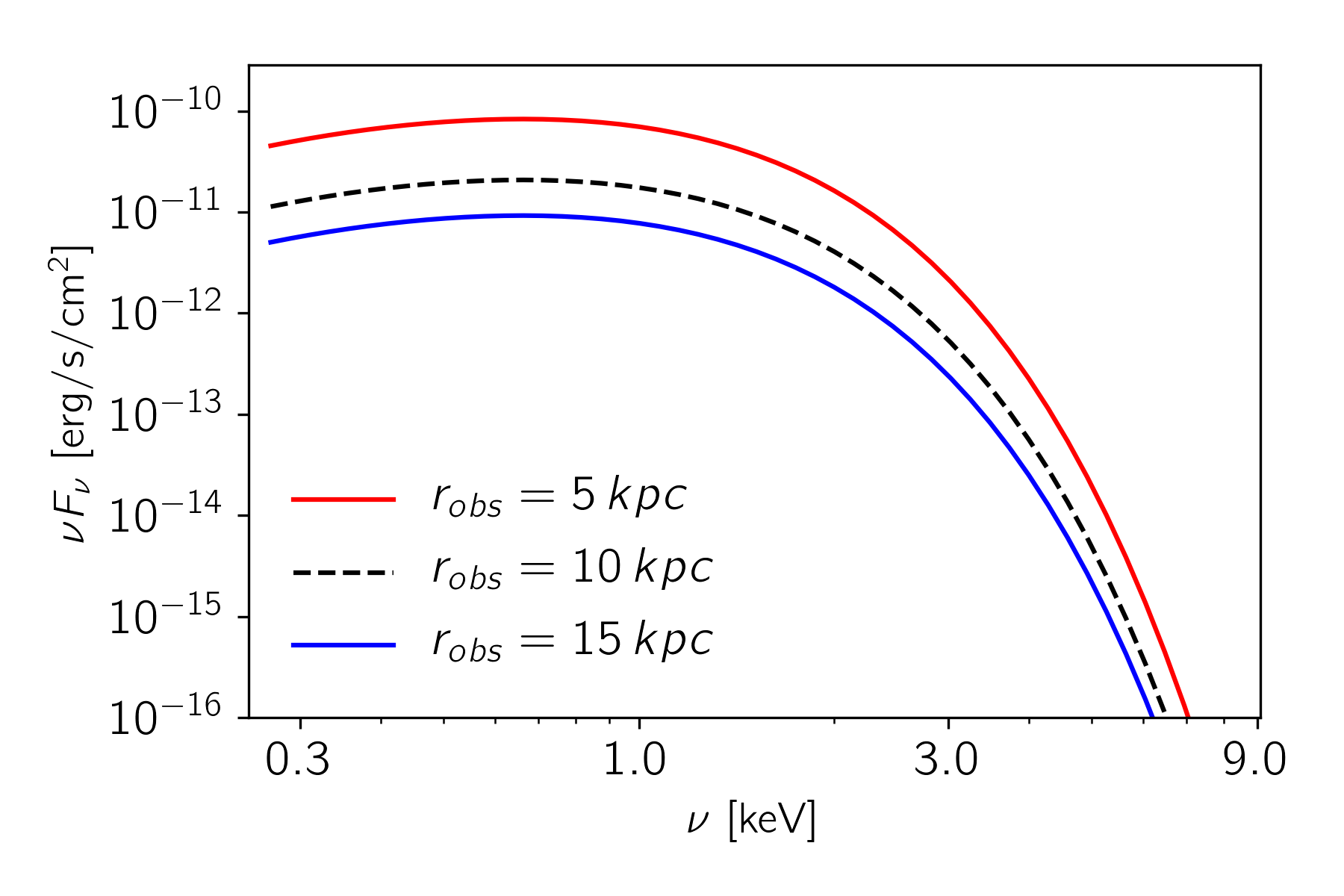} 
\includegraphics[width=.4\textwidth]{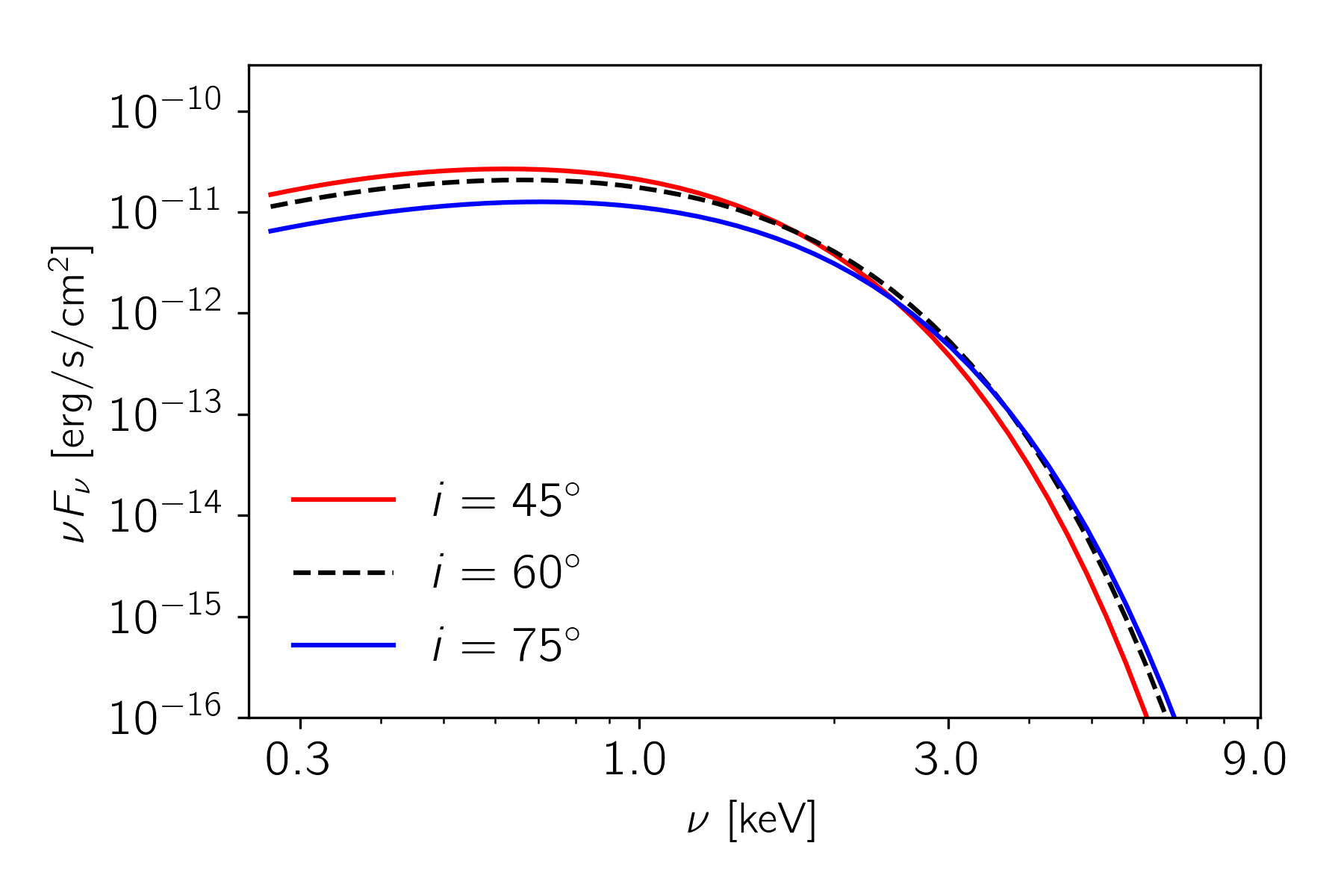} \\
\includegraphics[width=.4\textwidth]{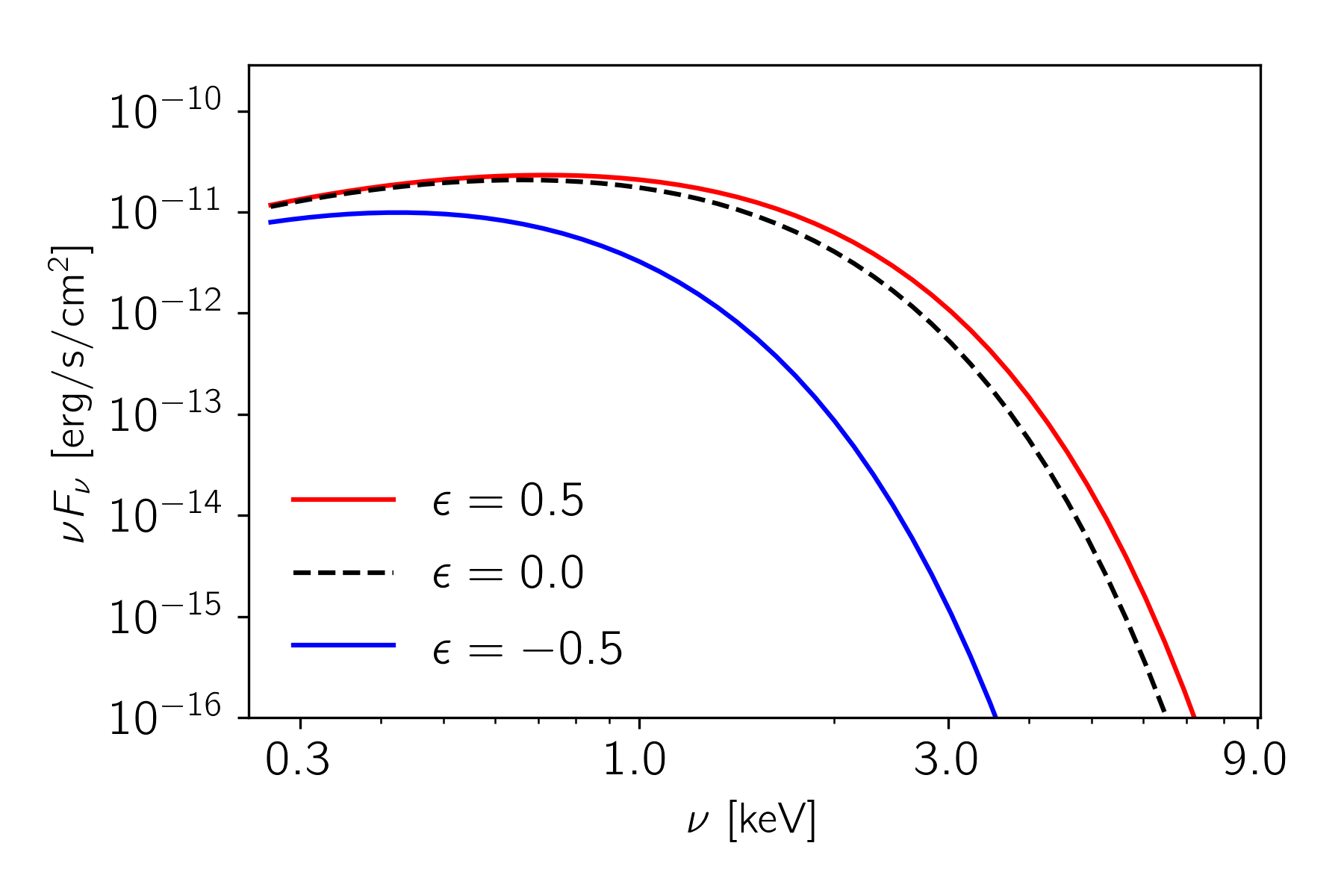}
\caption{(Color Online) Impact of the model parameters on the thermal spectrum of a thin disk. The parameters not specified in the caption are fixed to $\epsilon=0.0$, $\dot{M}=10^{18}\,\text{g/s}$, $M=10\,M_{\odot}$, $r_{\obs}=10\,\text{kpc}$ and $i=60^\circ$. Observe how a combination of a variation in $\dot{M}$ and in $i$ can mimic a variation in $\epsilon$.} 
\label{fig:SpectrumAll4}
\end{figure*}

The accretion disk spectrum is then fully determined by the parameters $\vec{\lambda} = (\epsilon, \dot{M}, M, r_{\obs}, i)$. Figure~\ref{fig:SpectrumAll4} shows how the spectrum is modified by changing one model parameter at a time. From this figure alone, it is already clear that parameter degeneracies are intrinsic to the model. For example, a combination of a simultaneous change in $\dot{M}$ and $i$ can mimic a simultaneous change in $i$ and $M$, as well as a change in $r_{\obs}$ and a change in $\epsilon$. Fortunately, however, some of these parameters, such as the source distance and the BH mass can be measured from other observations~\cite{remillard2006x,casares2006observational,unwin2008taking}, which can be used in our choice of priors to partially break some of these intrinsic degeneracies. 

%

\section{MCMC Methods}
\label{MCMC}

In this section, we describe the MCMC methods we use to explore the likelihood surface when carrying out parameter estimation.
 
Given a synthetic data injection $L_{\inj}$ characterized by $\bar{N}^{*}$ injected parameters $\vec{\lambda}^{*}=(\lambda^{*}_{1},\ldots,\lambda^{*}_{\bar{N}^{*}})$, we estimate the $\bar{N}$ parameters $\vec{\lambda}=(\lambda_{1},\ldots,\lambda_{\bar{N}})$ in the model $L_{\model}$ by minimizing the reduced $\chi^2$, defined by
\begin{equation}
\chi^2_{\rred}(\bar{N})=\frac{\chi^2}{F}=\frac{1}{F}\sum^F_{i=1}\left[\frac{L_{\model}(\nu_i,\vec{\lambda})-L_{\inj}(\nu_i, \vec{\lambda}^*)}{\sigma(\nu_i)}\right]^2,
\label{eq:Reducedchi}%
\end{equation}
where the summation is over $F=50$ sampling frequencies $\nu_i\in(10^{16.8},10^{18.3})\,[\text{Hz}]$ evenly spaced logarithmically. We model the standard deviation of the distribution, $\sigma$, via
\begin{equation}
\sigma(\nu_i)=\sum_{j}^{\bar{N}} \sigma_{j}(\nu_i),
\label{Sigma}
\end{equation}
where the sum is over the number of free parameters in the model, and where 
\begin{equation}
\sigma_{j}(\nu_i)=\frac{|L(\nu_i,\vec{\lambda}_{\notin n}^{*},\lambda^{*}_{n} + \delta \lambda^{*}_{n})
 - L(\nu_i,\vec{\lambda}_{\notin n}^{*},\lambda^{*}_{n} - \delta \lambda^{*}_{n})|}{2}.
\label{Sigma2}
\end{equation}
The term $\delta \lambda_n^{*}$ is a measure of the observational error in the injected parameter $\lambda_{n}^{*} \in \vec{\lambda}^{*}$. Based on previous works~\cite{remillard2006x,casares2006observational,reid2011trigonometric,narayan2013observational, casares2017x}, we here choose $\delta \epsilon=0.1$, $\delta \log\dot{M}=0.2\,\mathrm{g\,s^{-1}}$, $\delta M=1.0\,M_{\odot}$, $\delta r_{\obs}=2\,\mathrm{kpc}$ and $\delta i=5^\circ$.

We use the affine invariant MCMC sampler emcee~\cite{foreman2013emcee} to explore the likelihood surface and construct the posterior distribution of the model parameters. The latter is proportional to the the likelihood function 
\begin{equation}
\mathcal{L}(d|\vec{\lambda})= e^{- \chi^2_{\text{red}}(\vec{\lambda})/2}
\end{equation}
times the parameter priors. For the latter, we choose uninformative (flat) priors on the parameters $\epsilon$, $\dot{M}$ and $i$, with ranges $-0.9<\epsilon<0.9$, $17<\log\dot{M}\,[\mathrm{g\,s^{-1}}]<19$ and $0<i\,[\mathrm{rad}]<\pi/2$. For the parameters $r_{\obs}$ and $M$, we choose Gaussian priors with means $\mu_{r_{\obs}}=r_{\obs}^*$  $\mu_{M}=M^*$ and standard deviations $\sigma_{r_{\obs}}=\delta r_{\obs}$ and $\sigma_{M}=\delta M$. We make this choice because these parameters are typically known to some degree from independent observations~\cite{remillard2006x,unwin2008taking}. 

The ensemble of walkers is  always initialized by sampling from the prior distributions. For all the cases studied here, we burn-in the sampler for more than $50$ autocorrelation time scales and run until $\sim 100,000$ samples are obtained after burn-in. 
  
\section{Experimental Relativity Studies}
\label{Simulations}

In this section, we carry out different experimental relativity studies associated with accretion disk spectral observations. We classify the different studies based on whether the synthetic injection is constructed within GR or outside GR and whether the model used to recover the injection is built within GR or outside GR, as summarized in Table~\ref{tab:TypeSimulations}.  Cases A and B are those in which the injected signal is generated assuming GR, and we use either a GR (case A) or a non-GR model (case B), to recover the injection. Cases C and D are those in which the injected signal is generated with a non-GR model ($\epsilon \neq 0$), and we use either a GR (case C) or a non-GR model (case D) to recover the injection. In all cases, the synthetic injection and the model are both constructed as described in Sec.~\ref{Model}, so in the GR cases the parameter are $\vec{\lambda} = (\log\dot{M},M,r_{\obs},i)$ since $\epsilon = 0$, while in the non-GR cases the parameters are $\vec{\lambda} = (\epsilon, \log\dot{M},M,r_{\obs},i)$.

\begin{table}[t]
\begin{tabular}{|l|c|c|}\hline
\diaghead{\theadfont Diag ColumnmnHead II}%
  {Model}{Signal}&
\thead{GR}&\thead{non-GR}\\ \hline
GR & A & C \\    \hline
non-GR & B & D \\    \hline
\end{tabular}
\caption{Classification of cases studied in this paper. In case A, we inject a GR signal and attempt to extract it with a GR model, allowing us to estimate the accuracy to which accretion disk model parameters can be measured. In case B, we inject a GR signal and extract it with a non-GR model, allowing us to determine how well a non-GR deviation can be constrained. In case C, we inject a non-GR signal and extract with a GR model, allowing us to estimate the systematic uncertainties introduced in the extraction of accretion disk model parameters due to the a priori assumption that GR is correct. In case D, we inject a non-GR injection and extract with a non-GR model to determine whether GR deviations can be detected if they are present in the data.}
\label{tab:TypeSimulations}
\end{table}

The reason for these different studies is that each case allows a different type of investigation. Case A is perhaps the simplest, and in fact, what most of observed accretion disk spectra analysis implement. The goal of such a study would be to determine how well the parameters that describe the accretion disk model can be extracted given an observation. Case B is also a common investigation, whose goal is to determine how well a non-GR deviation can be constrained given an observation consistent with GR. Cases C and D, however, have not been studied as much. Case C allows one to determine how much systematic error one accrues by assuming GR is correct \emph{a priori} (an assumption sometimes referred to as \emph{fundamental bias}) if nature were to deviate from GR (see, for instance, Ref.~\cite{bambi2013measuring} for an example). Case D allows one to determine whether a GR deviation could be detected and differentiated from GR if nature were to deviate from GR. 

Even though we show results for mainly a few representative examples of certain combinations of injected parameters, the features we find are generic and based on an extensive numerical study in a very large region of parameter space. We present our results through corner plots that show the one and two dimensional projections of the posterior probability distributions of the parameters discussed in each case of study. The diagonal parts of the corner plots show the marginalized posterior distribution for each parameter, which allows one to read off the value of the best fit and the accuracy to which this best fit is determined. The off-diagonal parts of the corner plot show the two-dimensional projections of the likelihood, which show the covariances between parameters.

\subsection{Case A: Parameter Estimation in GR}

\begin{figure}[htb]
\includegraphics[width=\linewidth{}]{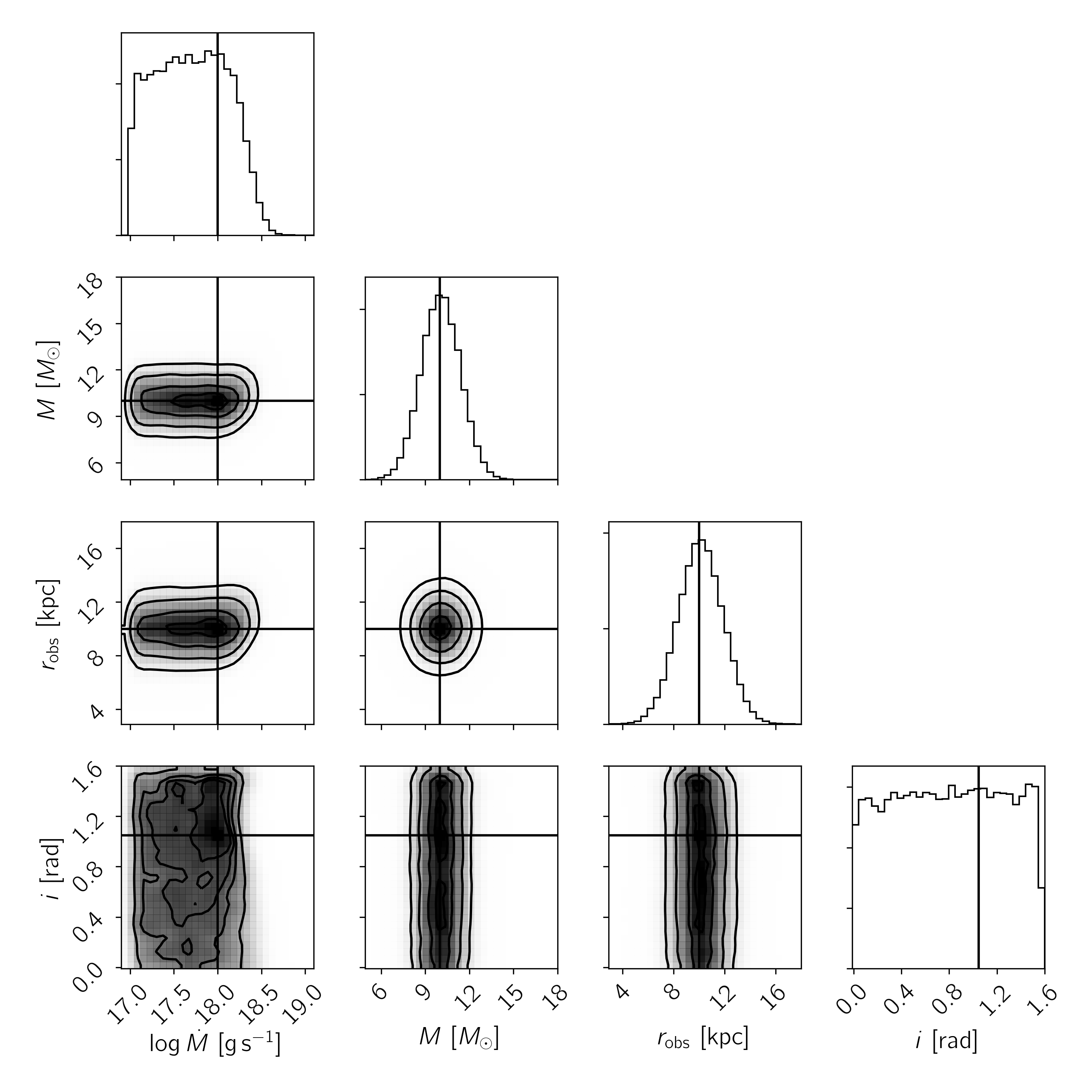}
\caption{Corner plot resulting from the MCMC analysis for case A. The vertical black lines correspond to the injected values. Observe that $\dot{M}$ and $r_{\obs}$ are well constrained, but this is because of the Gaussian priors we used. Observe also that $M$ and $i$ are much less well estimated due to the large degeneracies between these two parameters.} 
\label{fig:CornerGR}%
\end{figure}

We start with Case A: a GR injection extracted with the same GR model.  Figure~\ref{fig:CornerGR} shows the corner plot for this analysis. As expected, the posterior distribution of $M$ and $r_{\obs}$ are close to Gaussian, but this is not because the posterior is dominated by the likelihood, but rather it is because it is dominated by the Gaussian priors. The posterior distributions for $\dot{M}$ and $i$ tell a different story, with $\dot{M}$ measured to roughly one order of magnitude, while $i$ is not measured at all. This is because of the strong covariances between $\dot{M}$ and $i$ in the energy flux, which we already highlighted in Sec.~\ref{Model}. 

%

\subsection{Case B: Constraints on GR Deviations}

\begin{figure}[htb]
\includegraphics[width=\linewidth{},clip=true]{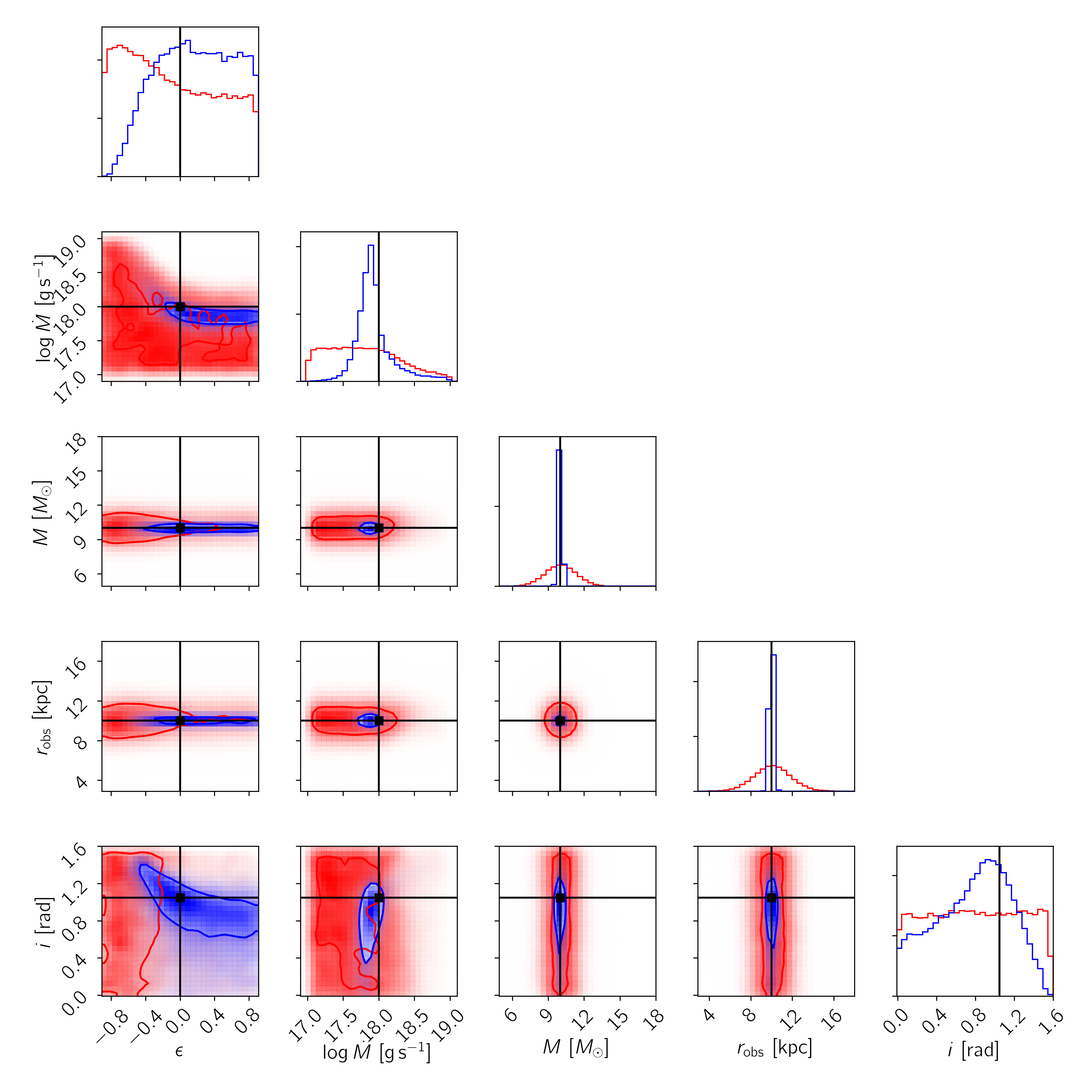}
\caption{(Color Online) Corner plot resulting from the MCMC analysis of case B, with the vertical black lines corresponding to the injected values. The red distributions correspond to the choice of observational error described below Eq.~\eqref{Sigma2} (consistent with current telescope capabilities), while the blue distributions correspond to observational errors that are one order of magnitude smaller (consistent with future telescope capabilities). Observe that the posterior on the bumpy parameter is consistent with the prior in both cases, with a slight improvement when decreasing the observational error.} 
\label{fig:CornerAll60}%
\end{figure}

We now move on to Case B: a GR injection recovered with a non-GR model. Figure~\ref{fig:CornerAll60} shows the corner plot for this analysis using the observational errors described below Eq.~\eqref{Sigma2} (red distributions), which are consistent with current telescope capabilities, as well as errors that are one order of magnitude smaller (blue distributions), which are consistent with future telescope capabilities. Observe that the posterior distribution on the bumpy parameter $\epsilon$ is nearly flat, changing very little from the initial flat prior distribution even when using future telescope capabilities. This is also the case for the parameters $\dot{M}$ and $i$ when using current telescope capabilities, which is consistent with the results of Fig.~\ref{fig:CornerGR}. However, when we employ future telescope capabilities, $\dot{M}$ and $i$ can now be measured.   

One way to quantify how similar or dissimilar the posterior distribution is relative to the prior distribution is through the Kullback-Leibler (KL) divergence measure~\cite{kullback1951information}. This quantity is defined by
\begin{align}
D_{n} = \int p_{r}(\lambda_{n}) \log_{2}\left[\frac{p_{r}(\lambda_{n})}{p(\lambda_{n})}\right] d\lambda_{n}\,,
\end{align}
where $p_{r}(\lambda_{n})$ is the prior on the $\lambda_{n}$, while $p(\lambda_{n})$ is the posterior. Clearly, when the posterior is equal to the prior, the divergence measure vanishes, while it increases logarithmically the more dissimilar these distributions are. The divergence measure can thus be thought of as quantifying the amount of information (measured in bits) that is gained by performing a given observation. Table~\ref{tab:KL} presents the divergence measure for case B, which confirms the conclusions of the previous paragraph: there is very little information gained in all parameters with current telescope capabilities. In particular, there is little information gain in $\epsilon$, but also in $M$ and $r_{\obs}$, because the posteriors of Fig.~\ref{fig:CornerAll60} are actually dominated by the Gaussian priors we chose. 

\begin{table}[t]
\begin{tabular}{|c|c|c|}
\hline 
 &  $D_{KL}\left[\sigma(\nu)\right]$ & $D_{KL}\left[\sigma(\nu)/10\right]$\tabularnewline
\hline 
\hline 
$\epsilon$ & $0.029$ & $0.45$\tabularnewline
\hline 
$\dot{M}$ & $0.186$ & $1.50$\tabularnewline
\hline 
$M$ & $0.037$ & $0.30$\tabularnewline
\hline 
$r_{\mbox{\tiny obs}}$ & $0.035$ & $0.32$\tabularnewline
\hline 
$i$ & $0.010$ & $0.25$\tabularnewline
\hline 
\end{tabular}
\caption{Information gain between prior to posterior in bits for the two cases shown in Fig.~\ref{fig:CornerAll60} for each parameter. The first column correspond to the choice of observational error described below Eq.~\eqref{Sigma2} (consistent with current telescope capabilities), while the second one correspond to observational errors that are one order of magnitude smaller (consistent with future telescope capabilities).}
\label{tab:KL}
\end{table}

\begin{figure}
\includegraphics[width=\linewidth{}]{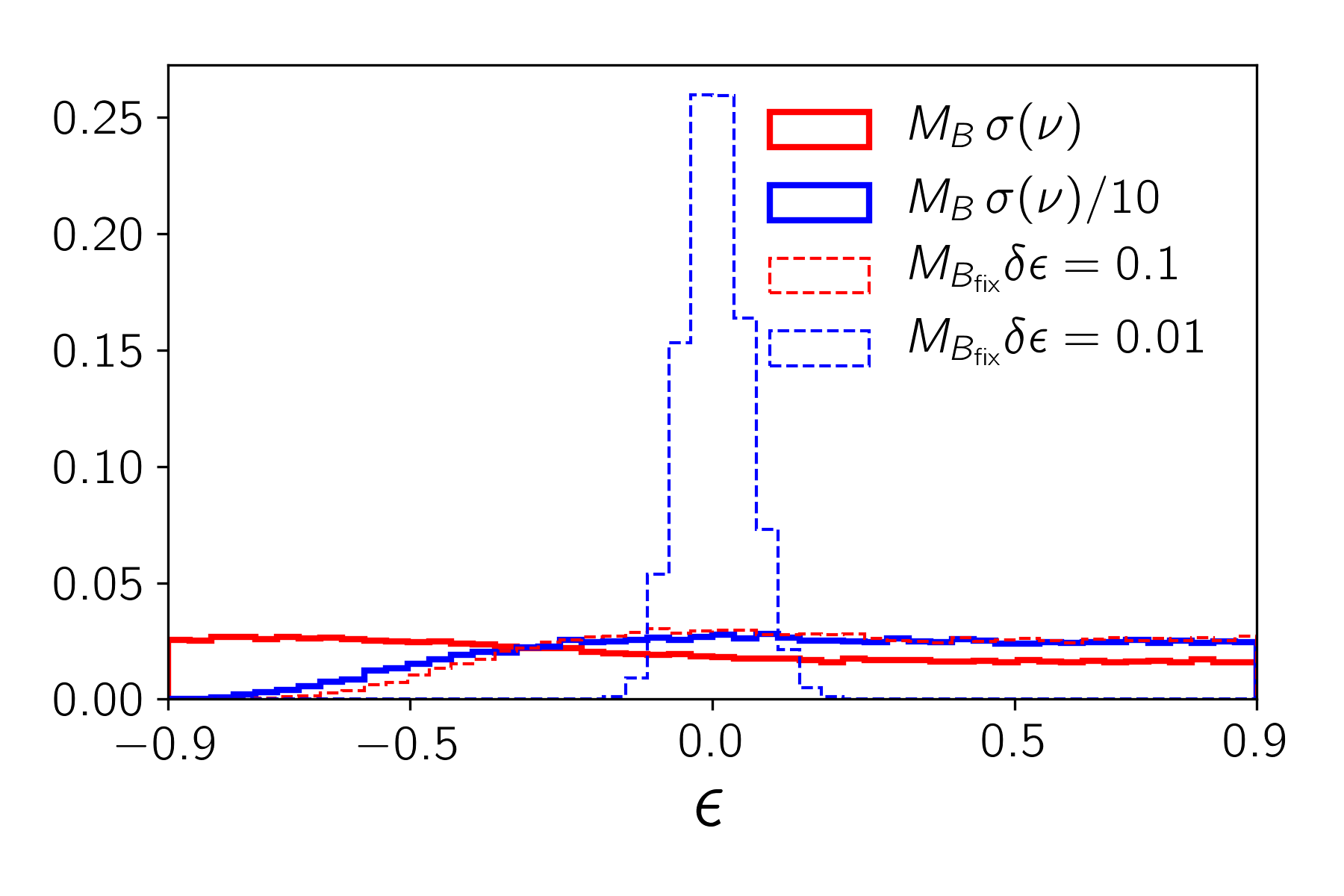}
\caption{(Color Online) Marginalized posterior distributions for the bumpy deviation parameter $\epsilon$ assuming current (red) and future (blue) telescope capabilities. The dashed distributions correspond to separate MCMC runs in which all parameters in the non-GR model are frozen to their injected values except for $\epsilon$. Observe how keeping the parameters of the model fixed could mislead us into believing we can constrain $\epsilon$ quite well, which is merely an artifact of ignoring the covariances in the model parameters.} 
\label{fig:RecoverEpsilon}%
\end{figure}

Let us now focus only on the posterior distribution of the bumpy deformation parameter $\epsilon$. Figure~\ref{fig:RecoverEpsilon} shows this posterior both assuming current telescope capabilities (red) and future telescope capabilities (blue). Moreover, this figure also shows a separate MCMC run in which we freeze all parameters in the non-GR model at the injected values, except for $\epsilon$ (dashed curves).  Observe that in general the posterior distributions are rather flat and consistent with the priors, indicating that one cannot constrain $\epsilon$ at all. However, if one freezes all parameters except for $\epsilon$ and uses a sufficiently small estimate for the accuracy to which $\epsilon$ can be measured, then one may be misled into believing that $\epsilon$ can be constrained rather well. This is clearly an artifact of freezing the parameters in the model, and therefore ignoring the important and strong covariances between the parameters, which tend to deteriorate our ability to test GR. Based on the results shown in Table~\ref{tab:KL}, an improvement of two orders of magnitude in telescope capabilities will provide a gain of information $D_{KL}$ of order unity, allowing $\epsilon$ to be constrained.

\subsection{Case C: Fundamental Bias}

\begin{figure}[htb]
\includegraphics[width=\linewidth{}]{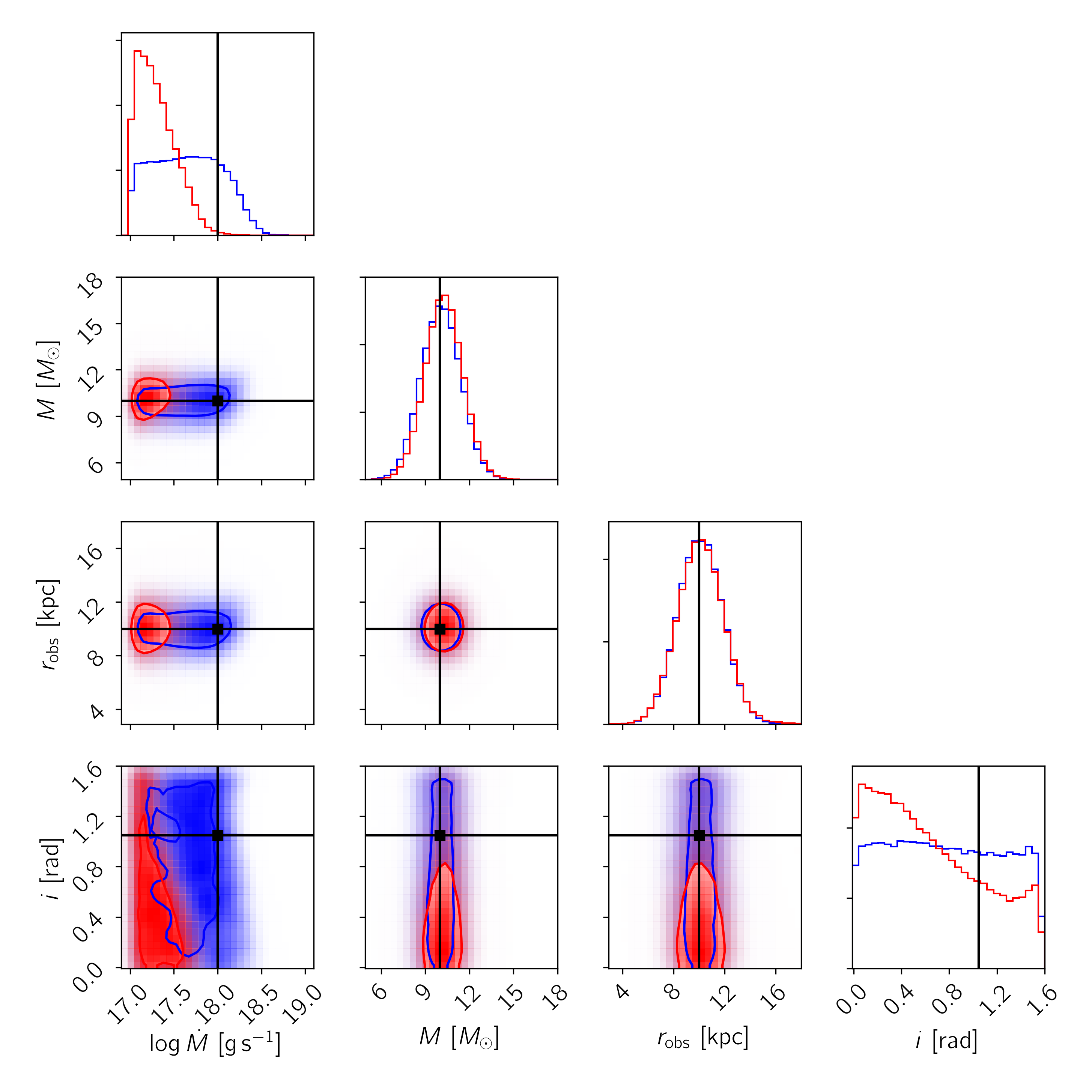}
\caption{(Color Online) Corner plot resulting from the MCMC analysis of case C. The vertical lines correspond to the injected values. The red distribution correspond to an injected bumpy deformation parameter of $\epsilon^{*} = -0.5$, while the blue distributions correspond to an injection with $\epsilon^{*} = -0.1$. Observe that when the GR deviation is negative enough, the posteriors for the accretion rate and the inclination angle peak significantly away from the injected value, indicating a systematic uncertainty due to fundamental bias.}
\label{fig:CornerFB}%
\end{figure}

Let us now consider case C: a non-GR injection extracted with a GR model. Figure~\ref{fig:CornerFB} shows the corner plot for two such analysis, where in one we injected a non-GR model with $\epsilon^{*}=-0.5$ (red) and in the other we injected $\epsilon^{*}=-0.1$ (blue). For the parameters $M$ and $r_{\obs}$ the effects of a non-GR signal are minimal because these parameters are already assumed to be well constrained independently through the Gaussian priors we chose. For the parameters $\dot{M}$ and $i$, however, the posteriors are quite different when we inject a non-GR deviation with a large enough bumpy parameter than what we found before. When $\epsilon^{*} = -0.1$ (blue case), the posteriors are indeed very similar to what we obtained with an $\epsilon^{*} = 0$ (GR) injection. But when $\epsilon^{*}  = -0.5$ (red case), the posteriors peak significantly away from the injected values. 

The conclusion of this analysis is that if astrophysical BHs are not described by the Schwarzschild  metric, but instead there is a sufficiently large deformation, then the extraction of accretion disk model parameters could be systematically biased. Care must be taken with this statement, however, since the value of the deformation that is required for this to happen is unrealistically large. Indeed, other observations using Solar System data, binary pulsar data, gravitational wave data or BH low-mass X-ray binaries have already constrained GR to a certain degree that typically would disallow values of $\epsilon$ as large as those injected here. As a particular example, the strongest constraint on Einstein-dilaton-Gauss-Bonnet gravity~\cite{Yunes:2011we} comes from low-mass X-ray binary observations~\cite{yagi2012new}. For this particular theory, the deformation parameter maps to $\epsilon=49/80 \left( 16 \pi \alpha^{2}/M^4\right)$, where $\alpha$ is the coupling constant of theory, implying that $\epsilon\lesssim 0.05$ for the systems studied here.

\subsection{Case D: Detecting Deviations from GR}

\begin{figure}[htb]
\includegraphics[width=\linewidth{}]{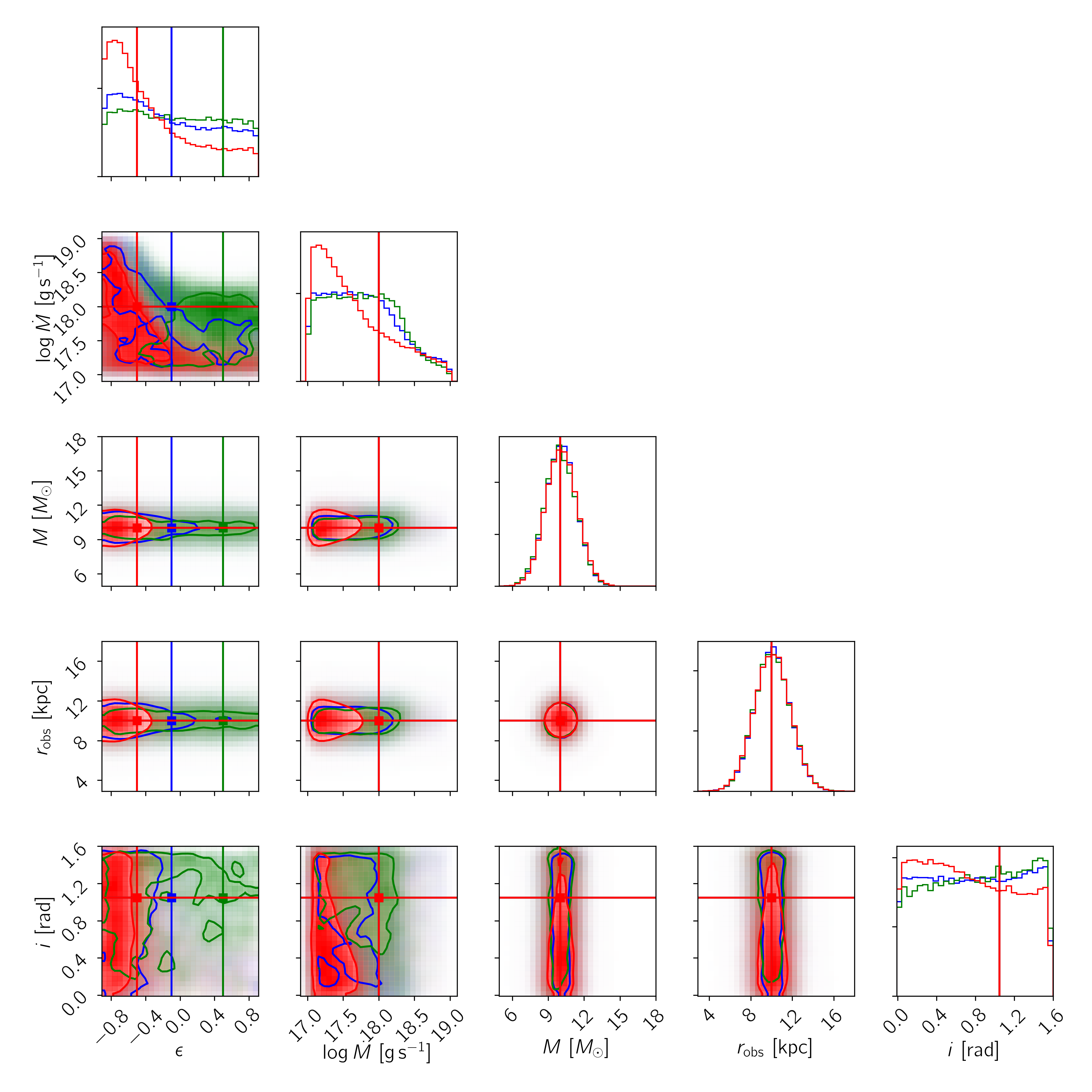}
\caption{(Color Online) Corner plot resulting from the MCMC analysis of case CD The vertical lines correspond to the injected values. The red distribution correspond to an injected bumpy deformation parameter of $\epsilon^{*} = -0.5$, the blue to $\epsilon^{*} = -0.1$ and the green to $\epsilon^{*} = +0.5$. Observe that when the GR deviation is negative enough, the posterior on $\epsilon^{*}$ peaks significantly away from zero, indicating the presence of a GR deviation. However, when the deformation is not large enough, then the accretion disk spectra are not sensitive enough to identify a deviation.} 
\label{fig:CornerD}%
\end{figure}

Finally, let us consider case D: a non-GR injection extracted with a non-GR model.  Figure~\ref{fig:CornerD} shows the corner plot for 3 such analysis, where we injected a non-GR signals with $\epsilon=-0.5$ (red), $\epsilon=-0.1$ (blue) and $\epsilon=+0.5$ (green). As the posteriors show, when the injected GR deviation is large enough, then the posterior on $\epsilon$ peaks significantly away from zero, indicating the presence of an anomaly that ought to be investigated further. However, for smaller deformations, as in the case of $\epsilon^{*} = -0.1$ and $\epsilon^{*} = +0.5$, the deformation is not significant enough to allow an observation of this type to distinguish between GR and non-GR. In this sense, one could thus easily be in a situation in which a GR deviation is present in the BH background, yet the observed accretion disk spectra is not sensitive enough to detect it (or distinguish a GR model from a non-GR model). 

One may wonder at this juncture why it is that an $\epsilon^{*} = -0.5$ allows for the identification of a non-GR deviation, while when $\epsilon^{*} = +0.5$ this is not the case. The answer is that when $\epsilon < 0$, the corrections introduced on the ISCO location are significantly larger than when $\epsilon>0$. Indeed, the ISCO radius can change to $12.33 M$ when $\epsilon^{*} = -0.5$, while it moves to $5.24M$ when $\epsilon^{*} = + 0.5$. Since the ISCO radius controls the location of the innermost edge of the accretion disk, this has a dominant effect on the magnitude of the non-GR correction to the flux.  

\section{Conclusions}
\label{Conclusions}

We simulated the thermal accretion disk spectra of stellar-mass BHs within and outside of GR, where the modifications were parameterized in terms of a bumpy deformation parameter. We have shown that the use of the accretion disk spectrum to constrain or identify non-GR deformations from a GR background is, at the very least, challenging. This is because of the degeneracy between the accretion disk model parameters and the non-GR bumpy parameter. Our analysis used a relatively simple spectral model, in which we ignored other sources of systematic error, such as those introduced due to calibration, other spectral components or incomplete modeling of the accretion disk physics. The inclusion of such effects can only strengthen the conclusions we arrived at, making it even more difficult to test GR with such observations. 

Our study also demonstrates the rather well-known fact (at least in some scientific communities) that carrying out a detailed exploration of the likelihood surface when estimating parameters is of crucial importance. Indeed, $\chi^{2}$ studies that focus on a particular sub-region of parameter space (e.g. freezing a subset of the model parameters) can greatly underestimate the effect of parameter covariances, and thus, be incorrectly led to too strong a set of conclusions on how well parameters can be measured. An MCMC exploration of the likelihood is a powerful technique to properly explore the likelihood, though it is computationally expensive and, as in the case studied here, it may require the use of high-performance computing clusters.   

Our work can of course be extended along several different directions. As a proof-of-study, our analysis focused on a rather simple accretion disk model, in which we ignored the BH spin, the accretion disk thickness, and other accretion disk physics (beyond those approximated in a simple geometrically thin/optically thick set-up). One could thus repeat our analysis to include these effects, and we expect that such a study will only strengthen our conclusions, i.e.~competitive tests of GR with accretion disk observations will be found to be even harder. This is because the inclusion of these effects does not increase the accuracy of the data set or produce any discernible features in the model, while it does increase the parameter degeneracies in the model. It is important to note that, while the effect on the deformation parameter studied here has a similar effect on the spectrum as the spin parameter, these two parameters are not completely degenerate, since spin also introduces frame dragging and other modifications to the metric that are different from those introduced by the $\epsilon$ parameter (though these differences are sub-dominant in the continuum spectrum). Therefore, in order to consider spin, one should deform away from the Kerr metric (not the Schwarzschild metric considered here), as otherwise one could be misled to believe one has measured a GR deviation when in reality this is not the case. That is why if a deviation is ever measured, the results should be taken with extreme caution.

Another possible direction for future research would be to investigate other non-GR deformations of the Kerr metric. In this work we focused on a particular bumpy metric, and on top of that, we restricted attention to a single deformation parameter. The inclusion of multiple parameters in the bumpy metric is necessary from a theory stand point, since no BH solution in any known modified theory leads only to a modification in the location of the event horizon, as we considered here. The inclusion of more parameters in the bumpy metric, however, will only increase the parameter degeneracies in the model, thus again worsening the constraints we can achieve on modified gravity. 

\acknowledgments

We gratefully acknowledge help and assistant from Frederic Vincent and Thibaut Paumard with the use and installation of GYOTO. We thank Erik Bryer for HPC guidance and Dimitry Ayzenberg, Cosimo Bambi, Nicholas Loutrel, Blake Moore, Remya Nair and Hector O. Silva for useful discussions and comments. A.C.-A. acknowledges funding from the Fundaci\'on Universitaria Konrad Lorenz (Project 5INV1). N.Y. and A.C.-A. acknowledge financial support through NSF CAREER grant PHY-1250636 and PHY-1759615, as well as NASA grants NNX16AB98G and 80NSSC17M0041. J.G. acknowledges financial support through the Montana Space Grant Consortium apprenticeship program. Computational efforts were performed on the Hyalite High-Performance Computing System, operated and supported by Montana State University's Information Technology Center.


\bibliography{References}

\end{document}